\documentstyle[epsfig]{mn}

\begin{document}

\def\fl{S_{\rm 1.4 \, GHz}}

\title[Measurement of the angular correlation function of radio galaxies from the NVSS]{Measurement of the angular correlation function of radio galaxies from the NRAO VLA Sky Survey}

\author[Chris Blake and Jasper Wall]{Chris Blake\footnotemark \, and Jasper Wall \\ Astrophysics, University of Oxford, Keble Road, Oxford, OX1 3RH}

\maketitle
\begin{abstract}
We quantify the angular distribution of radio sources in the NVSS by measuring the two-point angular correlation function $w(\theta)$.  By careful consideration of the resolution of radio galaxies into multiple components, we are able to determine both the galaxy angular clustering and the size distribution of giant radio galaxies.  The slope of the correlation function for radio galaxies agrees with that for other classes of galaxy, $\gamma = 1.8$, with a 3D correlation length $r_0 \sim 6$ h$^{-1}$ Mpc (under certain assumptions).  Calibration problems in the survey prevent clustering analysis below $\fl = 10$ mJy.  About 7 per cent of radio galaxies are resolved by NVSS into multiple components, with a power-law size distribution.  Our work calls into question previous analyses and interpretations of $w(\theta)$ from radio surveys.
\end{abstract}

\section{Introduction}
\renewcommand{\thefootnote}{\fnsymbol{footnote}}
\setcounter{footnote}{1}
\footnotetext{E-mail: cab@astro.ox.ac.uk}

Describing the large-scale structure of the Universe is of fundamental importance for testing theories of structure formation and measuring the cosmological parameters.  The angular distribution of galaxies, whilst merely a projection of the true 3D structure, is useful to quantify: it is easy to assemble a large sample of objects and it is possible to de-project the angular clustering (in a global statistical manner) to measure the 3D clustering.

Active galactic nuclei (AGN) detected at radio wavelengths are a powerful means of delineating large-scale structure.  They can be routinely detected over wide areas of the sky out to very large redshifts ($z \sim 4$) and hence describe the largest structures (and their evolution); radio emission is not sensitive to dust obscuration; accurate large-scale calibration should be possible; and the current generation of radio surveys such as WENSS, FIRST and NVSS contain AGN in very large numbers ($\sim 10^6$).

We quantify the angular distribution of AGN using the two-point angular correlation function \cite{pee}.  It is well-known that this method loses much of the clustering information: two very different structure morphologies can have the same correlation function.  However, the statistic provides a simple point of contact with prediction, the statistical errors are well-understood, it has a relatively simple de-projection into 3D, and perhaps most importantly, its measurement can reveal observational problems with the survey data.  It is an essential first analysis step before the application of more powerful techniques is warranted.

The key difference between angular correlation function analyses in the optical (e.g. Maddox, Efstathiou \& Sutherland 1996) and the radio regimes is in the latter, {\it the wide redshift range of radio sources washes out much of the clustering signal through the superposition of unrelated redshift slices.}  Hence an angular clustering signal was only marginally detected in datasets such as the 1.4 GHz Green Bank survey (Kooiman, Burns \& Klypin 1995) and the Parkes-MIT-NRAO survey (Loan, Wall \& Lahav 1997).  Deeper radio surveys such as FIRST (Becker, White \& Helfand 1995) and WENSS \cite{ren} revealed a clearer imprint of structure, quantified by the correlation function analyses of Cress et al. \shortcite{cre} and Magliocchetti et al. \shortcite{mag}.  Here we use the more extensive sky coverage and source list of the NVSS \cite{con} to test the robustness of previous conclusions with a higher signal-to-noise ratio.  We also reconsider the two critical observational effects: radio galaxies being resolved into separate components of radio emission, and calibration problems causing apparent source surface density gradients and discontinuities.

\section{Observational data and methods}

\subsection{The angular correlation function}

The {\it angular correlation function} $w(\theta)$ compares the observed (clustered) distribution to a random (unclustered) distribution of points across the same survey area, by simply measuring the fractional increase in the number of close pairs separated by angle $\theta$.  Specifically, if we let $DD(\theta)$ be the number of unique pairs of galaxies with separations $\theta \rightarrow \theta + \delta \theta$, and $RR(\theta)$ be the number of random pairs in the same separation range, then $w(\theta)$ can be estimated as
\begin{equation}
w_1 = \frac{DD}{RR} - 1
\label{eqw1}
\end{equation}
Landy \& Szalay \shortcite{lansza} point out that the estimator
\begin{equation}
w_2 = \frac{DD}{RR} - \frac{DR}{RR} + 1
\label{eqw2}
\end{equation}
which also involves the cross-pair count $DR(\theta)$, is superior because it has a much smaller variance, given by the `Poisson' error
\begin{equation}
\Delta w = \frac{1+w}{\sqrt{DD}}
\label{eqwerr}
\end{equation}
assuming the statistical error in the random sets can be neglected.  This can be achieved by averaging over a large number of random sets to obtain $\overline{DR}$ and $\overline{RR}$.

\subsection{The NVSS}

The 1.4 GHz NVSS (NRAO VLA Sky Survey, Condon et al. 1998) covers the sky north of declination $-40\degr$ (82 per cent of the celestial sphere).  The source catalogue contains $1.8 \times 10^6$ sources and is claimed to be 99 per cent complete at integrated flux density $\fl = 3.5$ mJy.  The survey was performed with the VLA in D configuration, with DnC configuration used for fields at high zenith angles ($\delta < -10\degr, \delta > 78\degr$), and the FWHM of the synthesized beam is 45 arcsec.  The raw fitted source parameters are processed by a publicly-available program NVSSlist, which performs the deconvolution and corrects for known biases to produce source diameters and integrated flux densities.  We used NVSSlist version 2.16 for our analysis.

Before measuring the angular correlation function from the survey, we must mask out various regions of the sky; identical masks must be used in the random comparison catalogues.  Firstly, we masked the survey within $5\degr$ of the Galactic plane to eliminate Galactic sources (such as supernova remnants).  Secondly, by virtue of the fitting algorithm, nearby bright extended radio galaxies can appear in the NVSS as a large number of separate elliptical Gaussians.  These objects introduce a very large number of spurious close pairs.  By visual inspection of the survey, we compiled a list of 22 masks around such galaxies.  Finally, the sidelobes of very bright sources may appear as spurious entries in the source catalogue.  As a precautionary measure, we placed circular masks of radius $0.5\degr$ around all radio sources brighter than 1 Jy (although ultimately this was found to have no effect on the measurement).

\subsection{Calibration/surface density problems}
\label{secsurf}

\begin{figure}
\epsfig{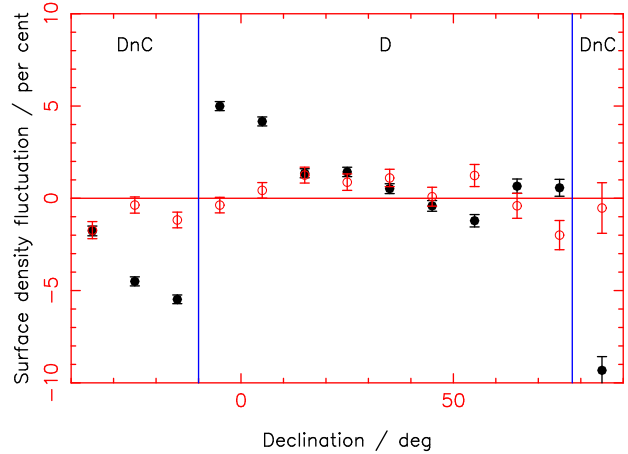}
\caption{Variations in NVSS source density as a function of declination for flux thresholds 2 mJy (filled circles) and 10 mJy (open circles).  The declination range of each array configuration is also indicated.  The error bar on the number of sources $N$ in a bin is $\sqrt{N}$.  Masked regions are excluded from this measurement.}
\label{figsurf}
\end{figure}

The NVSS suffers from calibration problems at low flux densities: spurious systematic fluctuations in source surface density.  As illustrated by Figure \ref{figsurf}, declination-dependent variations occur at flux densities below 10 mJy, including significant jumps at the declinations at which the array configuration changes.  These stem from the difficulty in compensating for the sparse {\it{uv-}}coverage of the NVSS (W.Cotton, private communication, 2001).

A varying source density will spuriously enhance the measured value of $w(\theta)$.  This is because the number of close pairs of galaxies depends on the local surface density ($DD \propto \overline{\sigma^2}$), but the number of close pairs in the random distribution depends on the global average surface density ($RR \propto (\overline{\sigma})^2$).  Systematic fluctuations mean $\overline{\sigma^2} > (\overline{\sigma})^2$, so according to equation \ref{eqw1}, $w(\theta)$ is increased.

To quantify this effect, we can show (Blake \& Wall, in preparation) that {\it on angular scales less than those on which $\sigma$ is varying}, $w(\theta)$ is subject to a constant offset $\overline{\delta^2}$, where $\delta = (\sigma - \overline{\sigma})/\overline{\sigma}$ is the surface over-density.  To estimate the magnitude of the offset, take a simple toy model in which a survey is divided into two equal areas between which there is an $\epsilon$ per cent shift in density.  A simple calculation shows $\overline{\delta^2} = \epsilon^2/4$.  So $\overline{\delta^2} \sim 2.5 \times 10^{-3}$ for NVSS sources above 2 mJy ($\epsilon \sim 0.1$), and $\overline{\delta^2} \sim 10^{-4}$ above 10 mJy ($\epsilon \sim 0.02$).

\section{Multiple-component sources}

The complex morphologies of radio sources mean that a single galaxy can be resolved in a radio survey as two or more closely-separated components of radio emission.  The median angular size of mJy radio sources is several arcseconds (Oort 1988).  Thus NVSS, with its beam-size of 45 arcsec, will leave the majority of this population unresolved.  However, there is a significant tail to the size distribution: many arcminute-size radio sources have been discovered (e.g. Lara et al. 2001) whose sub-structure will be resolved by NVSS.  These multiple components will produce spurious clustering at small separations.

Consider a single radio source in the survey.  How many pair separations would this contribute to a separation bin at angular distance $\theta$ of width $\delta \theta$?  This bin contains area $\delta A = 2 \pi \theta \, \delta \theta$ and hence the probability of another source falling in it is $\sigma_g \, \delta A \, [1 + w_g(\theta)]$ where $\sigma_g$ is the surface density of galaxies and $w_g$ is the galaxy angular correlation function.  Suppose another source does fall in the bin.  In general, both sources will be observed as a group of radio components: a single pair separation is replaced by several pair separations.  Some of these pair separations will lie {\it within an individual source}, the rest will be {\it between the different sources}.

Suppose $\theta$ is large enough that there are no pair separations of size $\theta$ within individual sources.  If there are on average $\overline{n}$ radio components per source, then there will be on average $(\overline{n})^2$ pair separations between these sources.  As the radio components will be roughly symmetrically distributed about their host galaxy, the average pair separation between the components remains equal to that of the original pair of sources, $\theta$.  It follows that the expected number of radio component pairs involving our original source in this separation bin is $\sigma_g \, \delta A \, [1 + w_g(\theta)] \times (\overline{n})^2$.  Hence over the whole survey, the expected number of radio component pairs in this bin is (neglecting edge effects)
\[ DD = \frac{1}{2} N_g \, \sigma_g \, \delta A \, [1 + w_g(\theta)] \, (\overline{n})^2 \]
where we have divided by 2 because all pairs are counted twice.  If $N_r$ and $\sigma_r$ are the total number and surface density of radio components, then $N_r = \overline{n} \times N_g$ and $\sigma_r = \overline{n} \times \sigma_g$ and this may be written
\[ DD = \frac{1}{2} N_r \, \sigma_r \, \delta A \, [1 + w_g(\theta)] \]
Generating a random set containing $N_r$ components, the number of random pairs in this bin is
\[ RR = \frac{1}{2} N_r \, \sigma_r \, \delta A \]
and hence the measured correlation function from the radio components (using equation \ref{eqw1}) is simply $w_r = w_g$.  So {\it multiple radio components have no effect on the measured correlation function at angular separations bigger than individual sources.}

At what separation $\theta$ may we neglect the effect of radio sources of size $\theta$?  The problem may be clarified by a realistic example.  Consider a separation bin at $\theta = 3$ arcmin = $0.05\degr$ of width $\delta \theta = 0.01\degr$.  Putting $\sigma = 10$ deg$^{-2}$, there are $\sigma 2 \pi \theta \delta \theta / 2 = 1.6 \times 10^{-2}$ random pairs {\it per original source} in this bin.  But we are trying to measure $w(\theta)$, which is a fractional enhancement $w_g(0.05\degr) \approx 2 \times 10^{-2}$ of this number of random pairs, which is $3 \times 10^{-4}$ pairs per original source.  Observations (e.g. Lara et al. 2001) indicate that $> 1$ in $10^4$ radio sources are as large as 3 arcmin.  Hence we cannot neglect pair separations within individual sources.  Thus {\it the small number of surplus pairs that determine the value of $w(\theta)$ mean that even a tiny fraction of giant radio sources can substantially change our measurement.}

Suppose $\theta$ is small enough that radio component pairs originating within individual sources are also important.  If $e$ is the fraction of sources observed to have multiple components, and $f(\theta) \, \delta \theta$ is the fraction of those component separations in the range $\theta \rightarrow \theta + \delta \theta$, then the total number of extra separations in the bin is $DD = N_g \, e \, f(\theta) \, \delta \theta$ and hence in this separation regime,
\begin{equation}
w_r(\theta) = w_g(\theta) + \frac{e \, f(\theta)}{(\overline{n})^2 \, \sigma_g \, \pi \, \theta}
\label{eqmult}
\end{equation}

The angular scale on which the dominant source of NVSS pairs changes from multi-component sources to individual sources is 6 arcmin ($0.1\degr$), as evidenced by a clear break in the measured correlation function (see Figure \ref{figcorr}).

\section{Results}

\subsection{The angular correlation function}
\label{secres}

\begin{figure}
\epsfig{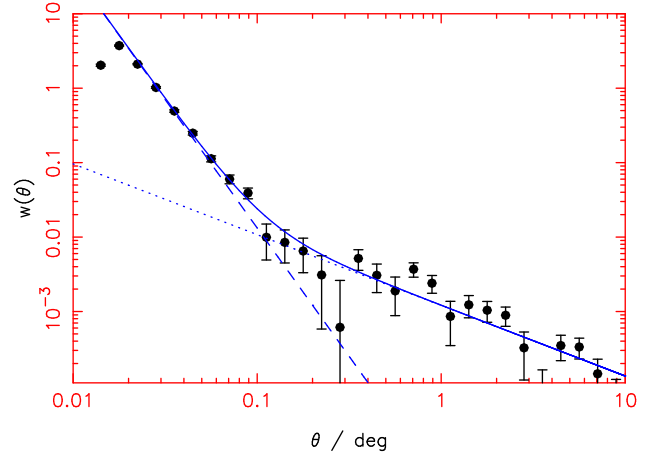}
\caption{Measurement of $w(\theta)$ for NVSS sources with $\fl > 15$ mJy.  The best-fitting sum of two power-laws is overplotted.}
\label{figcorr}
\end{figure}

\begin{figure}
\epsfig{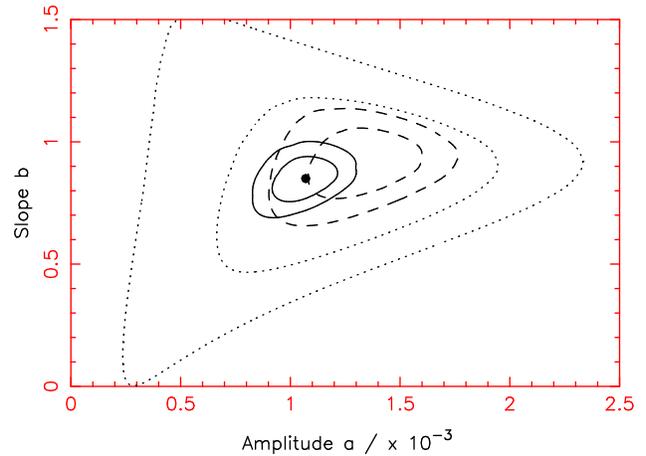}
\caption{Contours of constant $\chi^2$ in the space of the clustering parameters $(a,b)$.  $1\sigma$ and $2\sigma$ contours are shown ($\chi^2$ increasing by 2.30 and 6.17 from its minimum) for flux thresholds 50 mJy (dotted), 20 mJy (dashed) and 10 mJy (solid).  The bold circle indicates the best-fitting combination for the 10 mJy threshold.}
\label{figparam}
\end{figure}

\begin{table*}
\caption{Best-fitting amplitudes of a sum of two power-laws, for a range of flux thresholds ($\theta$ is measured in degrees).  The $1\sigma$ ranges of the best-fitting parameters and the best-fitting reduced $\chi^2$ are also displayed.  $N$ is the number of sources analyzed at each threshold.}
\label{tabres}
\begin{tabular}{ccccccc}
\hline
Flux threshold & $N$ & Clustering power-law & 1$\sigma$ & Size power-law & 1$\sigma$ & $\chi_{red}^2$ \\
/ mJy & & $(a \times 10^{-3}) \, \theta^{-0.8}$ & range & $(c \times 10^{-5}) \, \theta^{-3.4}$ & range & \\
\hline
50 & 114,362 & $a = 1.25$ & $0.87 \rightarrow 1.63$ & $c = 1.52$ & $1.45 \rightarrow 1.60$ & 1.55 \\
30 & 192,610 & $a = 1.33$ & $1.10 \rightarrow 1.55$ & $c = 0.94$ & $0.90 \rightarrow 0.98$ & 1.41 \\
20 & 281,514 & $a = 1.26$ & $1.11 \rightarrow 1.42$ & $c = 0.68$ & $0.65 \rightarrow 0.71$ & 1.36 \\
15 & 361,644 & $a = 1.12$ & $0.99 \rightarrow 1.24$ & $c = 0.58$ & $0.55 \rightarrow 0.60$ & 1.78 \\
10 & 504,045 & $a = 1.04$ & $0.95 \rightarrow 1.13$ & $c = 0.42$ & $0.40 \rightarrow 0.44$ & 2.30 \\
5 & 838,740 & $a = 1.29$ & $1.23 \rightarrow 1.34$ & $c = 0.24$ & $0.23 \rightarrow 0.25$ & 4.37 \\
\hline
\end{tabular}
\end{table*}

Figure \ref{figcorr} displays the measured $w(\theta)$ for $\fl > 15$ mJy.  We use the Landy-Szalay estimator (equation \ref{eqw2}) with the error from equation \ref{eqwerr}.  A good fit to $w(\theta)$ is a sum of two power-laws.  One power-law dominates at small angles ($\theta < 0.1\degr$) and is a direct indication of {\it the size distribution $f(\theta)$ of giant radio sources}.  The other power-law dominates at large angles ($\theta > 0.1\degr$) and describes {\it the clustering between individual radio sources.}  We measured $w(\theta)$ for a variety of flux thresholds and fitted a sum of two power-laws to the results.  Turning first to radio galaxy clustering, we find:

\begin{itemize}
\item The slope of the clustering power-law does not depend on flux threshold and is consistent with $w(\theta) = a \, \theta^{-0.8}$, as found for optically-selected galaxies.  Our most accurate determination (for a 10 mJy threshold) is $- 0.85 \pm 0.1$.
\item The clustering amplitude is also independent of flux threshold: our most accurate value is $a = 1.04 \pm 0.09 \times 10^{-3}$ (with $\theta$ in degrees).
\item Below $\fl \sim 10$ mJy, the calibration problems described in section \ref{secsurf} become dominant, as evidenced by a sudden increase in the $\chi^2$ of the best fit, as well as direct measurement of the surface density (Figure \ref{figsurf}).
\end{itemize}

To view the errors on the clustering parameters $w(\theta) = a \, \theta^{-b}$, in Figure \ref{figparam} we plot contours of constant $\chi^2$ in $(a,b)$ parameter space for flux thresholds 50 mJy, 20 mJy and 10 mJy.  As the survey becomes deeper, the increasing number of sources enables us to measure the clustering parameters more accurately until the calibration problems intervene.

Considering the size distribution, we find:

\begin{itemize}
\item The slope of the small-angle correlation function is independent of flux threshold and has the value $-3.4$.
\item The amplitude of the small-angle correlation function changes with flux threshold (equivalently surface density $\sigma$) as $1/\sigma$, as predicted by equation \ref{eqmult}.
\end{itemize}

In table \ref{tabres} we list the best-fitting amplitudes for both power-laws for the flux thresholds considered, obtained by minimizing the $\chi^2$ statistic.  For purposes of comparison, we fix the slopes of the small-angle and large-angle power-laws at $-3.4$ and $-0.8$ respectively.  The fits are performed to angles $\theta > 2$ arcmin $= 0.033\degr$, safely above the resolution limit of the NVSS.  We derive $1\sigma$ errors on the fitted amplitudes by varying each in turn from the best-fitting combination and finding when $\chi^2$ increases by 1.0 from its minimum.  The reduced $\chi^2$ statistic of the best fit is also indicated.

Previous correlation function analyses of radio surveys (Cress et al. 1996, Magliocchetti et al. 1998) did not consider either the gradients in source surface density (present in FIRST as well as NVSS, Blake \& Wall in preparation) or the large angular scales on which multi-component sources affect the measured $w(\theta)$.  Conclusions from these analyses must be regarded as suspect.

\subsection{Determination of spatial clustering properties}

\begin{figure}
\epsfig{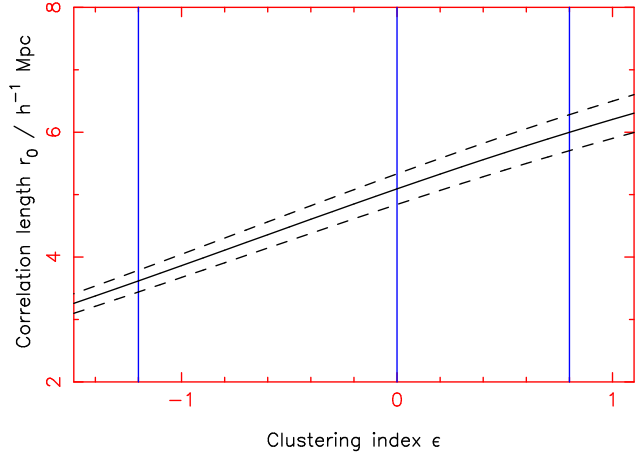}
\caption{Determination of the spatial clustering length $r_0$ of NVSS sources for a range of clustering indices.  The solid line corresponds to the best-fitting angular clustering amplitude and the dashed lines encompass the $1\sigma$ range.  The vertical lines mark the special values of $\epsilon$ mentioned in the text.}
\label{figr0ee}
\end{figure}

\begin{figure}
\epsfig{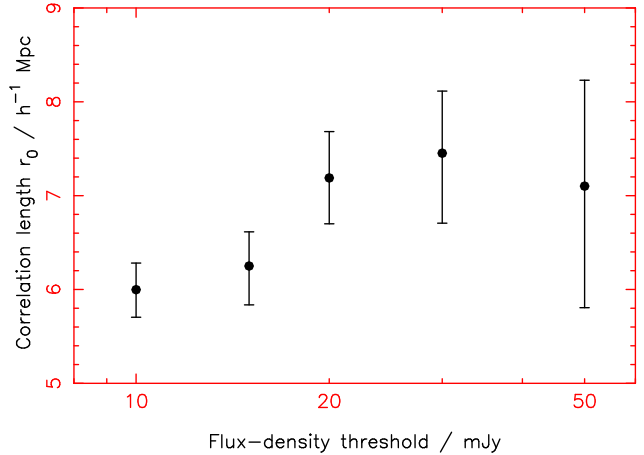}
\caption{Determination of the clustering length $r_0$ of NVSS sources for a range of flux-density thresholds, assuming a clustering index $\epsilon = \gamma - 1 = 0.8$.}
\label{figr0fl}
\end{figure}

The angular distribution of galaxies is a projection of the true 3D distribution.  Not knowing the individual redshifts of the NVSS sources, we cannot infer their full spatial distribution.  However, we can infer their spatial clustering properties from their angular clustering knowing only their {\it redshift distribution} $N(z)$ \cite{loa}.  This is not surprising: clustering properties are global statistical measures of the sample, like the redshift distribution.

The spatial correlation function $\xi(r)$ is usually parameterized by a power-law of slope $\gamma$ and `correlation length' $r_0$.  To allow for redshift evolution of clustering we introduce the `clustering index' $\epsilon$ such that
\[ \xi(r) = \left( \frac{r}{r_0} \right)^{-\gamma} (1 + z)^{-3-\epsilon} \]
$\epsilon = 0$ implies constant clustering in proper co-ordinates, $\epsilon = \gamma-3$ implies constant clustering in co-moving co-ordinates, and $\epsilon = \gamma-1$ represents growth of clustering under linear theory (Peebles 1980).

It can be shown \cite{pee} that a power-law $\xi(r)$ projects on to the celestial sphere as another power-law $w(\theta) = a \, \theta^{1-\gamma}$, where the projection involves an integral over $N(z)$.  Conversely, knowing the amplitude and slope of $w(\theta)$, we can deduce $\gamma$ and $r_0$.  This is discussed further by Cress \& Kamionkowski \shortcite{crekam} and Magliocchetti et al. \shortcite{mag}.  Our measured slope of $-0.8$ demonstrates that $\gamma = 1.8$; to deduce $r_0$ we must also assume a value for $\epsilon$, a form for $N(z)$ and a cosmology (we take $\Omega = 1$, $\Lambda = 0$).  In the current absence of observed redshift distributions for complete samples of mJy radio galaxies, we use the $N(z)$ predicted from the luminosity-function models of Dunlop \& Peacock \shortcite{dunpea}.  With these assumptions, in Figure \ref{figr0ee} we use our 10 mJy measurement of $w(\theta)$ to deduce $r_0$ for a range of $\epsilon$.

In Figure \ref{figr0fl}, we fix $\epsilon = \gamma-1$ and investigate if $r_0$ depends on flux threshold (using the appropriate $N(z)$ for each threshold).  The result is $r_0 \sim 6$ h$^{-1}$ Mpc with a marginal dependence on flux.  A more detailed study of spatial clustering inferences will be the subject of a future paper.

\subsection{Determination of the size distribution}
\label{secsize}

Our measurements of the small-angle $w(\theta)$ indicate that the size distribution of giant radio galaxies ($\theta > 2$ arcmin) is a power-law, $f(\theta) \propto \theta^{-2.4}$, at all flux thresholds considered.  (This differs from the correlation function slope $-3.4$ by virtue of the extra power of $\theta$ in equation \ref{eqmult}).  Such a steep slope is naturally produced in toy models: if radio sources have linear sizes up to a maximum $L_0$, then the available volume for sources with angular sizes $> \theta$ scales as $\theta^{-3}$.

\begin{figure}
\epsfig{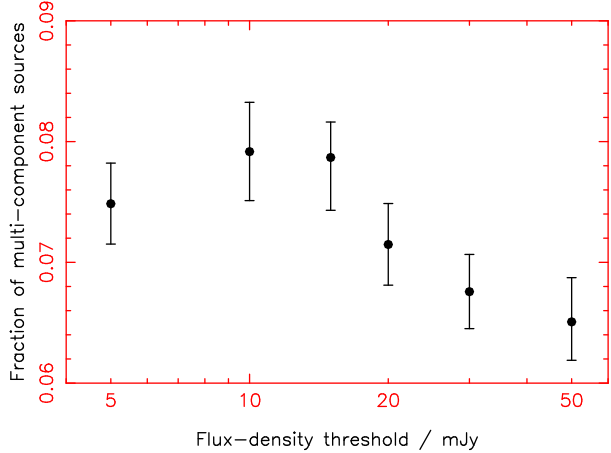}
\caption{The fraction of radio galaxies that are resolved by NVSS into multiple components, as a function of flux threshold.}
\label{figfrac}
\end{figure}

We can use the measured amplitude of the small-angle $w(\theta)$, in conjunction with equation \ref{eqmult}, to determine the fraction $e$ of radio galaxies that are resolved by NVSS into multi-component sources.  Figure \ref{figfrac} illustrates that $e \sim 0.07$.

To compare these numbers with observations, we note that Lara et al. \shortcite{lar} used visual inspection of the NVSS to find $\sim 80$ radio galaxies with sizes greater than $\theta_0 = 4$ arcmin and total fluxes $\fl > 100$ mJy in the region $\delta > 60\degr$.  Integrating our derived size distribution above $\theta_0$ predicts a total of 62 such objects.

\section{Conclusions}

Using the NVSS radio survey, we have measured the angular correlation function $w(\theta)$ of radio galaxies with unprecedented precision.  The results may be summarized as follows:

\begin{enumerate}
\item The correlation function has two contributions: that due to multiple components of the same galaxy, dominant at $\theta < 0.1\degr$, and that due to clustering between galaxies, which dominates at larger angles.  A clear break in $w(\theta)$ is evident between these scales.
\item The clustering part has a slope consistent with that measured for other classes of galaxy, $w(\theta) \propto \theta^{-0.8}$.
\item The clustering amplitude corresponds to a spatial clustering length $r_0 \sim 6$ h$^{-1}$ Mpc (under certain assumptions), independent of flux-density threshold.
\item The NVSS suffers from calibration problems that prevent the measurement of $w(\theta)$ at flux densities below 10 mJy.
\item The size distribution of arcminute radio sources is well-described by a power-law with slope $-2.4$; $\sim 7$ per cent of galaxies are resolved by NVSS into multiple components.
\end{enumerate}

For the first time, our investigation has untangled the imprint of radio galaxy clustering from the other observational effects, in particular the resolution of radio galaxies into multiple components.  As such it opens a new observational window for large-scale structure investigations, as well as providing a novel means of measuring the size distribution of giant radio galaxies.

\section*{acknowledgments}
We thank Lance Miller and Steve Rawlings for very helpful comments on earlier drafts of this paper.

\end{document}